\documentclass[paper]{JHEP3} 

\usepackage{epsfig,multicol,bbm,citesort}

\newcommand\fverb{\setbox\fverbbox=\hbox\bgroup\verb}
\newcommand\fverbdo{\egroup\medskip\noindent%
			\fbox{\unhbox\fverbbox}\ }
\newcommand\fverbit{\egroup\item[\fbox{\unhbox\fverbbox}]}
\newbox\fverbbox
\newcommand\Z{\mathbb{Z}}
\newcommand\cl{\clearpage}

\preprint {}

\title{Revisiting the deconfinement phase transition in $SU(4)$ Yang-Mills 
theory in 2+1 dimensions}

\author{
Kieran Holland\\
Department of Physics, University of the Pacific\\
3601 Pacific Ave, Stockton CA 95211, USA\\
E-mail: \email{kholland@pacific.edu}}

\author{
Michele Pepe\\
INFN, Istituto Nazionale di Fisica Nucleare, Sezione di Milano-Bicocca\\
Edificio U2, Piazza della Scienza, 3 - 20126 Milano, Italy\\
E-mail: \email{pepe@mib.infn.it}}

\author{
Uwe-Jens Wiese\\
Institute for Theoretical Physics, Bern University\\
Sidlerstrasse 5, CH-3012 Bern, Switzerland \\
E-mail: \email{wiese@itp.unibe.ch}}

\abstract{In order to deepen our understanding of the nature of the 
deconfinement phase transition for various gauge groups, we
investigate $SU(4)$ Yang-Mills theory in $2+1$ dimensions. We find
that the transition is weakly first order. We perform extensive Monte
Carlo simulations on lattices with temporal extent $N_t = 3, 4,$ and
5, and spatial sizes up to $N_s = 20\; N_t$. We observe coexistence
of confined and deconfined phases at the critical temperature, and
finite-size scaling shows consistency with first order exponents. The
continuum extrapolation of the latent heat yields $L_h/T_c^3=0.188(17)$.} 


\begin{document} 

\section{Motivation}

The confinement of quarks inside baryons and mesons is a feature of the
strong interactions only at low temperatures. At sufficiently high temperatures
hadrons melt and quarks and gluons form a plasma. Numerical simulations of QCD 
support the expectation that this process takes place in a smooth manner and 
the high- and low-temperature regimes are analytically connected through a 
crossover. As one increases or decreases the quark masses, the situation 
changes and, at some point, a finite temperature phase transition occurs. 
This phase transition is related to symmetries that are either badly broken or 
only approximate when the quarks have their physical masses. For massless 
quarks,
chiral symmetry is exact: it is spontaneously broken at low temperatures and it
gets restored in a chiral phase transition at finite temperature. In the 
opposite limit, when the quarks are heavy, they are only weakly coupled to the 
gluons. As the quark mass becomes much larger than the typical energy scale of 
the strong interactions, $\Lambda_{\rm QCD}\approx 250$~MeV, the quarks decouple
and gluons are the only relevant degrees of freedom. In the limit of infinitely
heavy quarks, the global center symmetry of Yang-Mills theory is no longer 
explicitly broken by the quarks' triality and it becomes an exact symmetry of
the theory. When the temperature is about 300~MeV, the center symmetry breaks
spontaneously at a phase 
transition~\cite{Polyakov:1978vu,Susskind:1979up,Holland:2000uj}. Since quarks 
transform non-trivially under center transformations, the breaking of the 
center symmetry implies deconfinement.

Yang-Mills theory provides a simplified framework in which the phenomenon of
confinement can be investigated without facing the more difficult numerical 
problems related to dynamical fermions. Moreover, the pure gluon dynamics is 
quite rich and
many investigations have been performed to study various interesting features 
of Yang-Mills theory. For instance, string effects in the static quark 
potential have been observed numerically~\cite{Kut05}. The study of topological
objects and of their relevance for the mechanism of confinement is an
active field of investigation~\cite{Gre03}. A systematic study of $SU(N)$ 
Yang-Mills theory for various $N$ is a research topic that aims at 
understanding the way in which the large $N$ limit is 
approached~\cite{tHo73,Wit79}. This paper deals with another important 
characteristic of Yang-Mills theory, namely the order of the deconfinement 
phase transition.

About 25 years ago, Svetitsky and Yaffe conjectured~\cite{Sve82} that the 
critical behavior of a gauge theory at the deconfinement transition can be 
described by a scalar field theory with a symmetry corresponding to the center 
of the gauge group. In fact, if one integrates out the spatial components of 
the gluon field in $(d+1)$-dimensional Yang-Mills theory, one obtains an 
effective action for the scalar field represented by the Polyakov loop. The
corresponding scalar field theory is defined in $d$ dimensions and, 
in general, its action is very complicated. However, if the deconfinement phase
transition happens to be second order, as one approaches the critical point, 
the correlation length diverges and universal critical behavior arises. Hence, 
the details of the complicated effective action become irrelevant: only the 
center symmetry and the dimensionality of space determine the universality 
class.

Svetitsky and Yaffe's conjecture has been checked in many numerical simulations
in Yang-Mills theory with various gauge groups, both in $2+1$ and in $3+1$ 
dimensions. In those cases in which the deconfinement phase transition is 
second order, the universality class has indeed turned out to be the one
predicted by Svetitsky and Yaffe. In $3+1$ dimensions, $SU(2)$ Yang-Mills 
theory has a second order deconfinement phase
transition~\cite{Kut80,McL81,McL81a,Eng81,Gav83,Gav83a} in the universality 
class of the 3-dimensional Ising model~\cite{Eng89,Eng92}. However, in $3+1$ 
dimensions no other pure gauge theory has been found to have a second order 
deconfinement phase 
transition~\cite{Cel83,Kog83,Got85,Bro88,Fuk89,Alv90,Win00,Luc02,Luc04,Luc05}.
In $2+1$ dimensions, $SU(2)$ Yang-Mills theory again has a second order 
deconfinement phase transition, now in the universality class of the 
2-dimensional Ising model~\cite{Tep93,Eng97}. Since in $2+1$ dimensions 
fluctuations are stronger than in $3+1$ dimensions, there are two more 
cases in which the deconfinement phase transition is second order. At its
deconfinement phase transition, $(2+1)$-dimensional $SU(3)$ Yang-Mills theory
shows the same critical behavior as the 2-dimensional 3-state Potts
model~\cite{Eng97,Chr92,Chr92a}. The group $Sp(2)$ has the same center $\Z(2)$
as $SU(2)$, and the deconfinement phase transition of the corresponding 
$(2+1)$-d $Sp(2)$ Yang-Mills theory is again in the universality 
class of the 2-dimensional Ising model~\cite{Hol03}. For other gauge groups, 
the deconfinement phase transition of the corresponding pure gauge theory has 
turned out to be of first order~\cite{Hol05,Lid05}. 

In this paper, we examine $SU(4)$ Yang-Mills theory in $2+1$ dimensions 
because it is one of the last remaining unsettled cases. The original 
study on coarse lattices indicated that the transition is second 
order~\cite{Gross:1984pq}. The improved numerical results presented 
in~\cite{deF03} show that, on coarse lattices, the transition is weakly first
order. For finer lattices, the deconfinement phase transition appeared to be 
second order, perhaps belonging to the universality class of the 2-dimensional
4-state Potts model. This is a particular case of the 2-d $\Z(4)$-symmetric 
Ashkin-Teller model which has lines of critical points along which the 
universality class and the critical exponents change continuously. The authors 
of~\cite{deF03} pointed out that it is difficult to obtain a definite answer 
unless one considers rather large volumes and they could not rule out a weak 
first order phase transition in the continuum limit. The need for
large volumes and fine lattices was also emphasized in~\cite{Lid05}.

In this paper we present numerical evidence for a weak first order 
deconfinement phase transition in $SU(4)$ Yang-Mills theory in $2+1$ 
dimensions. Interestingly, although there is an infinite set of different 
available universality classes, the system deconfines with non-universal 
behavior. The center symmetry does not play a role in determining the order of 
the deconfinement phase transition. Only when the transition is second
order does the center symmetry determine the universality class. As we
conjectured in~\cite{Hol03}, the order of the deconfinement phase transition is
determined by the size of the group. In the low-temperature confined phase, the
dynamics of Yang-Mills theory is governed by glueballs. The number of 
glueball states --- i.e.\ the number of singlets in the tensor product 
decomposition of adjoint representations --- is essentially independent of the 
gauge group. On the other hand, the dynamics of the high-temperature plasma
phase is determined by deconfined gluons, whose number is given by the number
of generators of the gauge group. If there is a large mismatch between the 
number of relevant degrees of freedom in the confined and the deconfined 
phases, the phase transition does not proceed smoothly as a second order 
transition. Instead an abrupt discontinuous first order transition takes place.
This conjecture is supported by numerical simulations which show that the 
strength of the first order transition increases with the size of the gauge 
group. Further evidence was provided by studies of Yang-Mills theory with the
exceptional gauge group $G(2)$~\cite{Hol03a}. The group $G(2)$ is the smallest,
simply connected group with a trivial center. Therefore, in $G(2)$ Yang-Mills 
theory there is no symmetry argument that implies the presence of a 
finite temperature deconfinement phase transition. However, since $G(2)$ (which
has 14 generators) has a rather large size, $G(2)$ Yang-Mills theory has a 
first order deconfinement phase transition in $3+1$ 
dimensions~\cite{Pep05,Pep06,Cos07}. Various aspects concerning the problem of
confinement in $G(2)$ Yang-Mills theory have been investigated
in~\cite{Gre06,Maa07}. 

The rest of the paper is organized as follows. In section 2, we describe the 
standard lattice formulation of Yang-Mills theory, the observables that we
consider, and the finite-size scaling analysis used to determine the order of
the phase transition. The numerical results are presented in section 3, 
followed by our conclusions.

\section{$SU(4)$ Yang-Mills Theory on the Lattice}

\subsection{The action and the observables}

We perform numerical simulations of $SU(4)$ Yang-Mills theory on a periodic 
lattice in $2+1$ dimensions. We consider the standard Wilson plaquette action
\begin{equation}
S[U] = - \frac{\beta}{4} \sum_\Box \mbox{Re} \mbox{Tr} \ U_\Box =
- \frac{\beta}{4} \sum_{x,\mu <\nu} \mbox{Re} \mbox{Tr} \
(U_{x,\mu} U_{x+\hat\mu,\nu} U^\dagger_{x+\hat\nu,\mu} U^\dagger_{x,\nu}),
\end{equation}
where the link parallel transporter matrices $U_{x,\mu} \in SU(4)$ are group 
elements in the fundamental representation. All dimensionful quantities are
expressed in units of the lattice spacing. The bare dimensionful gauge 
coupling $\beta$ is related to the usual gauge coupling $g$ in the continuum by
$\beta = 8/g^2$. The path integral measure and the partition function $Z$ then 
take the form 
\begin{equation}
\int {\cal D}U = \prod_{x,\mu} \int_{SU(4)} dU_{x,\mu},
\hspace{1.5cm}
Z = \int {\cal D}U \exp(- S[U]).
\end{equation}
The Polyakov loop~\cite{Polyakov:1978vu,Susskind:1979up}
\begin{equation}
\Phi_{\vec x} = \frac{1}{4} \mbox{Tr}({\cal P} \prod_{t = 1}^{N_t} 
U_{\vec x,t,d+1})
\end{equation}
is the trace of a path-ordered product of link variables along a loop wrapping 
around the periodic Euclidean time direction. Here $N_t = 1/T$ is the extent of
the lattice in the Euclidean time direction, which determines the temperature 
$T$ in lattice units. The expectation value of the Polyakov loop is given by
\begin{equation}
\langle \Phi \rangle = 
\frac{1}{Z} \int {\cal D}U \ \frac{1}{N_s^2} \sum_{\vec x} \Phi_{\vec x} 
\exp(- S[U]),
\end{equation}
where $N_s$ is the extension of the lattice in the spatial directions. The 
Polyakov loop represents a scalar field that transforms non-trivially under
symmetry transformations in the center subgroup $\Z(4)$ of $SU(4)$. Hence, a
non-vanishing expectation value of the Polyakov loop indicates the spontaneous
breakdown of the center symmetry and thus signals deconfinement. However, in a
finite periodic volume spontaneous symmetry breaking --- in the sense of a
non-vanishing order parameter --- cannot occur. Therefore, in the finite-size 
scaling analysis discussed below, we will consider the expectation value of the
magnitude of the Polyakov loop $\langle |\Phi| \rangle$. In a finite volume 
this quantity is always non-vanishing but it approaches zero when one takes the
thermodynamic limit in the confined phase. Another quantity that is useful for 
distinguishing the confined from the deconfined phase is the probability 
distribution for the Polyakov loop,
\begin{equation}
p(\Phi) = \frac{1}{Z} \int {\cal D}U \ \delta\left(\Phi - \frac{1}{N_s^2} 
\sum_{\vec x} \frac{1}{4} \mbox{Tr}({\cal P} \prod_{t = 1}^{N_t} 
U_{\vec x,t,d+1})\right) \exp(- S[U]).
\end{equation}
In the confined phase $p(\Phi)$ has a single peak centered at $\Phi = 0$. In 
the deconfined phase, on the other hand, it has four degenerate maxima at 
$\Phi = \Phi_0 \exp(i k \pi/2)$, where $\Phi_0$ is a positive real number and 
$k = 0,1,2,3$. When the deconfinement phase transition is first order, the 
confined and the deconfined phases coexist and can be distinguished by their
different values of the Polyakov loop even at the phase transition. In that 
case, close to the phase transition one thus observes five maxima of the
distribution $p(\Phi)$. The relative weight of the confined and deconfined 
peaks changes as one crosses the phase transition. On the other hand, when 
the deconfinement phase transition is second order, the high- and
low-temperature phases become indistinguishable at criticality. The
confined maximum becomes broader and broader as the critical
temperature is approached from below and,  at criticality, the width
of the peak diverges. When the temperature is  increased further, the
four $\Z(4)$-symmetric deconfined peaks emerge smoothly from the broad
distribution of the Polyakov loop. In the limit of very high
temperatures, one can perform an analytic perturbative calculation of
the effective potential for the Polyakov
loop~\cite{Wei80,Wei81,Bel89,Pis06,Hoy07}. 

Other useful observables that characterize the deconfinement phase transition
are the Polyakov loop susceptibility $\chi$, defined by
\begin{equation}
\chi= N_s^2 \left( \langle |\Phi|^2 \rangle - \langle |\Phi| \rangle^2 \right),
\end{equation}
and the specific heat $C$ given by
\begin{equation}
C = \frac{1}{3N_s^2 N_t}\left(\langle S^2 \rangle - \langle S \rangle^2 \right).
\end{equation}
For a first order transition it is interesting to also consider the latent heat
$L_h$. As we said above, in the case of a first order transition one can 
distinguish the confined from the deconfined phases even at the transition. One
can then define the action densities $s_c$ and $s_d$ for the confined and the 
deconfined phases, respectively. The latent heat is defined as the difference 
in the action density between the two phases at the critical temperature and
it is given by
\begin{equation}
L_h = s_d - s_c.
\end{equation}
The fluctuations in the action attain their maximum when the two phases have 
the same probability. It then follows that, in the thermodynamic limit, the
maximum $C^{\rm max}$ of the specific heat is given by
\begin{equation}
C^{\rm max} = \frac{3N_s^2 N_t}{4} L_h^2.
\end{equation}
For a second order deconfinement phase transition the latent heat vanishes
since the confined and deconfined phases become indistinguishable at the 
critical point.

\subsection{Finite-size scaling}

Away from a phase transition, the susceptibility of an extensive quantity 
scales with the volume. This scaling behavior changes as we approach a phase 
transition and the fluctuations become stronger. For a first order transition, 
the susceptibility of an extensive quantity increases with the square of the 
volume. This scaling behavior follows from the coexistence of the two phases. 
In fact, in general, an observable has different values in the two phases. 
Since the observable is extensive, its susceptibility scales with the square of
the volume. In case of a second order phase transition, the susceptibility
scales faster than the volume but --- unlike for a first order transition ---
not as fast as the square of the volume. The exponent that characterizes the 
scaling behavior depends on the observable and on the universality class of the
phase transition. 

The method we have used in the finite-size scaling analysis of our numerical
data is the following. For a given lattice size $N_s^2\times N_t$, we perform a
set of numerical simulations at various couplings $\beta$ across the 
deconfinement phase transition. Using the Ferrenberg-Swendsen re-weighting 
technique~\cite{Fer88,Fer89} we determine a pseudo-critical coupling 
$\beta_{c,N_s,N_t}$ from the maximum of the Polyakov loop susceptibility. We
then repeat this procedure for various values of the spatial lattice size 
$N_s$. For a first order phase transition the critical coupling depends on the
spatial volume as
\begin{equation}
\beta_{c,N_s,N_t} = \beta_{c,N_t} + a_0 \frac{N_t^2}{N_s^2} + \ldots
\label{eq:beta_c}
\end{equation}
where $\beta_{c,N_t}$ is the critical coupling in the limit of infinite spatial
volume at fixed temporal extent $N_t$. 

Up to corrections to scaling, the data for the Polyakov loop susceptibility 
density $\chi/N_s^2$ collected at different couplings $\beta$ and for different 
lattice sizes $N_s^2\times N_t$ collapse onto a single universal curve once they 
are plotted as a function of the finite-size scaling variable 
$x=(N_s/N_t)^2 \ (\beta/\beta_{c,N_s,N_t}-1)$. The corrections to scaling can 
then be easily measured at the maximum of the curve
\begin{equation}
\frac{\chi^{\rm max}}{N_s^2} = \left(\frac{\chi^{\rm max}}{N_s^2}\right)_{\infty}
+ b_0 \frac{N_t^2}{N_s^2} + \ldots
\label{eq:chi}
\end{equation}
A similar formula holds for the maximum of the specific heat
\begin{equation}
\frac{C^{\rm max}}{3 N_s^2 N_t} = \frac{1}{4} L_h^2 + c_0
\frac{N_t^2}{N_s^2} + \ldots
\label{eq:latent}
\end{equation}

\section{Discussion of the numerical results}

\subsection{Simulation details}
We have performed simulations on lattices with $N_t=3,4$, and 5, and for 
spatial sizes $N_s$ as large as $20\; N_t$. We have used a standard
combination of heat-bath~\cite{Creutz:1980zw} and 
over-relaxation~\cite{Adler:1981sn,Adler:1987ce,Creutz:1987xi,Brown:1987rra}
algorithms to update the various $SU(2)$ subgroups of
$SU(4)$~\cite{Cabibbo:1982zn}. We have simulated with a ratio of
over-relaxation to heat-bath updates of 4/1 and 1/1, and we find no
significant difference between these two choices. For each set of $\beta, N_t$,
and $N_s$, we have generated at least $10^5$ configurations to be used for 
measurements. These runs are sufficiently long such that for the
smaller physical volumes with e.g.~$N_s/N_t=10$, we see
$\cal{O}$(50-100) tunneling events between the various bulk
phases. For the larger volumes like $N_s/N_t=20$, this is reduced to
$\cal{O}$(10) tunnelings. We expect that this is a reasonable
sampling of the various bulk phases.

\subsection{Monte Carlo histories and Polyakov loop distributions}

In Figure~\ref{fig:mc_history}, we plot the Monte Carlo histories of the 
Polyakov loop and the plaquette expectation value, configuration by 
configuration, for a $60^2 \times 3$ lattice at $\beta=20.40$. The system 
spends a long time in a particular phase, characterized by the value of $\Phi$,
before it rapidly tunnels to a different phase, in which it again remains for a
significant period of Monte Carlo time. In this particular run, we see the 
confined phase, in which $\Phi$ fluctuates around 0, and four deconfined 
phases, in which $\Phi$ varies around the four values $\Phi_0 \exp(i k \pi/2)$. 
The lower plot shows that the plaquette changes simultaneously with the 
Polyakov loop, between two similar but still distinguishable values. This 
suggests that the system is close to the phase transition, and that the
deconfinement transition is first order, with coexisting confined and
deconfined phases. However, the small jump in the plaquette value indicates 
that the transition may well be rather weak. This simulation is quite
typical in that we see about 10 tunneling events occur in this large
volume. 

\TABLE[h]{
\begin{tabular}{ccc} 
\hline \hline
$N_t$ & $N_s$ & $\beta_{c,N_s,N_t}$ \\
\hline \hline
3 &  26 & 20.251(8) \\
  &  28 & 20.271(6) \\
  &  30 & 20.288(7) \\
  &  36 & 20.3230(25) \\
  &  42 & 20.351(5) \\
  &  48 & 20.363(4) \\
  &  60 & 20.390(4) \\
  &  72 & 20.388(5) \\
\hline 
4 &  32 & 25.966(14) \\
  &  36 & 26.019(14) \\
  &  40 & 26.065(9) \\
  &  48 & 26.097(10) \\
  &  52 & 26.135(12) \\
  &  56 & 26.173(10) \\
  &  64 & 26.203(11) \\
  &  80 & 26.192(9) \\
\hline 
5 &  34 & 31.64(4) \\
  &  40 & 31.75(4) \\
  &  46 & 31.72(4) \\
  &  50 & 31.77(3) \\
  &  56 & 31.91(4) \\
  &  60 & 31.977(26) \\
  &  66 & 32.012(24) \\
  & 100 & 32.090(22) \\
\hline \hline
\end{tabular}
\caption{The finite-volume critical couplings for various values
of $N_s$ and $N_t$, with bootstrap errors.}
\label{table:beta_c_NsNt}
}

In Figure~\ref{fig:poly_dist}, we show the probability distributions of the 
complex-valued Polyakov loop for simulations on $36^2 \times 3$ lattices at 
three different $\beta$ values, ranging from low to high temperature. At low 
temperature, there is just the confined phase, while at high
temperature there are four  deconfined phases. At $\beta = 20.26$, the
system is apparently quite close to the transition temperature and the
five bulk phases coexist. Figure~\ref{fig:plaq_dist} shows the
probability distributions of the plaquette value for the same
simulations. We find only a single-peak distribution, which varies
smoothly with $\beta$. In the case of a normal-strength first 
order transition, close to the critical temperature one would expect to see two
distinct peaks. In the present case, the discontinuity in the plaquette value
is clearly visible in the Monte Carlo history --- as shown in the bottom part 
of Figure~\ref{fig:mc_history} --- but due to its small size, it does not stand
out in the plot of the probability distribution.

\subsection{Polyakov loop susceptibility}
We use re-weighting of the various ensembles to determine the location 
$\beta_{c,N_s,N_t}$ of the peak of the Polyakov loop susceptibility $\chi$. This
is an accurate method and we show typical results in
Figures~\ref{fig:reweight_42s_3t} and \ref{fig:reweight_48s_4t}. We
use the bootstrap method to calculate both the error in $\chi$ for
each individual ensemble, and the error in $\beta_{c,N_s,N_t}$
extracted from the re-weighting of several ensembles. We list the
finite-volume critical couplings in Table~\ref{table:beta_c_NsNt} for
the various simulations we have performed.

If the deconfinement transition is of first order, the infinite-volume
critical coupling should be approached asymptotically as
$(N_t/N_s)^2$, as described in Equation~(\ref{eq:beta_c}). The data as
shown in Figures~\ref{fig:beta_crit_3t}, \ref{fig:beta_crit_4t}, and
\ref{fig:beta_crit_5t} display exactly this behavior. In determining
the infinite-volume critical coupling for the various $N_t$ values,
there is some systematic error involved in the choice of the
extrapolation range. One does not know how large $N_s$ has to be
before linear behavior sets in. Let us first discuss the $N_t=3$
data. We start by fitting all of the data, then omit one at a time the
data for the smaller values of $N_s$. The results are listed in
Table~\ref{table:beta_c_sys}, with a statistical error included for each
fit. For $N_t=3$, we see that the fit does not improve as data is
omitted, so the optimal fit includes all of them. For the final error
estimate, we use the jackknife method applied to the optimal data
set. We quote a final result of $\beta_{c,N_t}=20.414(5)$. The optimal
fit and this extrapolated value are also shown in
Figure~\ref{fig:beta_crit_3t}. 

\TABLE[h]{
\begin{tabular}{cccc} 
\hline \hline
$N_t$ & $N_{\rm data}$ &$\beta_{c,N_t}$ & $\chi^2$/d.o.f. \\
\hline \hline
3 & 8 & 20.414(4) & 6.9/6 \\
  & 7 & 20.415(4) & 6.2/5 \\
  & 6 & 20.416(5) & 5.5/4 \\
  & 5 & 20.418(6) & 4.8/3 \\
  & 4 & 20.416(10) & 4.6/2 \\
  & 3 & 20.417(18) & 4.5/1 \\
final value & & 20.414(5) & \\
\hline 
4 & 8 & 26.251(13) & 17.4/6 \\
  & 7 & 26.253(16) & 17.0/5 \\
  & 6 & 26.253(20) & 17.0/4 \\
  & 5 & 26.26(3) & 16.2/3 \\
  & 4 & 26.24(3) & 8.0/2 \\
  & 3 & 26.22(3) & 2.8/1 \\
final value & & 26.251(16) & \\
\hline 
5 & 8 & 32.14(4) & 26.1/6 \\
  & 7 & 32.18(4) & 16.2/5 \\
  & 6 & 32.22(4) & 7.8/4 \\
  & 5 & 32.21(5) & 7.4/3 \\
  & 4 & 32.164(21) & 0.69/2 \\
  & 3 & 32.153(5) & 0.02/1 \\
final value & & 32.22(8) & \\
\hline \hline
\end{tabular}
\caption{The infinite-volume critical couplings for the various values
of $N_t$ and the quality of the linear extrapolations, where some of the
smaller $N_s$ data are discarded.}
\label{table:beta_c_sys}
}

We apply exactly the same procedure to the $N_t=4$ results. From
Figure~\ref{fig:beta_crit_4t}, we see that a linear fit of all of the
data gives $\beta_{c,N_t}=26.251(13)$, the error being statistical
only. However, the quality of the fit is poor, with $\chi^2$/d.o.f.\ =
17.4/6. Omitting the data for smaller $N_s$, the quality of the fit
does not improve, in fact it becomes worse, as listed in
Table~\ref{table:beta_c_sys}. The small $N_s$ data do not seem to be
at fault, as significant fluctuations occur at larger $N_s$. A
quadratic fit gives $\beta_{c,N_t}=26.26(3)$ and $\chi^2$/d.o.f.\ =
17.2/5, so the data show almost no quadratic behavior and the
extrapolated value agrees very well with the linear fit. Hence, the
optimal choice seems to be to use all of the data. Performing a jackknife
error analysis on this set, the final result we quote is
$\beta_{c,N_t}=26.251(16)$. 

For both $N_t=3$ and 4, a linear extrapolation appears valid for a
fitting range $(N_t/N_s)^2 \le 0.015$. We expect that the same should hold
for $N_t=5$. Looking at the data in Figure~\ref{fig:beta_crit_5t},
two of the data lie outside this range. At sufficiently small $N_s$,
one does expect to see deviations from linearity. A linear fit of all of the
data gives a very poor quality of fit. When the two smallest $N_s$ values
are omitted, $\chi^2$/d.o.f.\ improves significantly, from 26.1/6 to
7.8/4, giving $\beta_{c,N_t}=32.22(4)$, the error being statistical
only. As listed in Table~\ref{table:beta_c_sys}, the fit improves
further if the next two small $N_s$ data are discarded. However, we
believe that this is excessive and gives the impression of a more
accurate extrapolation than is warranted. Using the linear regimes of
$N_t=3$ and 4 as a guide, we conclude that the optimal fit excludes
only the two smallest $N_s$ data. For comparison, we also fit all of the data
using a quadratic form, which gives $\beta_{c,N_t}=32.24(6)$ and
$\chi^2$/d.o.f.\ = 14.9/5. The extrapolated value is completely
consistent with that from a linear fit, although the quadratic fit is
somewhat inferior. Using the jackknife method on the optimal set, we
quote a final value for $N_t=5$ of $\beta_{c,N_t}=32.22(8)$. 

We also investigate the finite-size scaling behavior of the Polyakov
loop susceptibility.
In Figures~\ref{fig:uni_sus_3t}, \ref{fig:uni_sus_4t}, and
\ref{fig:uni_sus_5t}, we plot the rescaled susceptibility 
$\chi/N_s^2$ as a function of the finite-size scaling variable, 
$x=(N_s/N_t)^2 \ (\beta/\beta_{c,N_s,N_t}-1)$, for $N_t = 3,4$, and 5.
We are assuming here that the exponents are those of a first 
order transition. We see that, close to the critical temperature, the
data do indeed collapse onto a single universal curve. This is further
evidence of the first order nature of the deconfinement transition.

\TABLE[h]{
\begin{tabular}{ccc} 
\hline \hline
$N_t$ & $N_s$ & $C^{\rm max}/(3N_s^2N_t) \times 10^4$ \\
\hline \hline
3 &  24 & 13.66(7) \\
  &  26 & 11.81(6) \\
  &  28 & 10.22(6) \\
  &  30 & 8.95(6) \\
  &  36 & 6.46(4) \\
  &  42 & 4.89(3) \\
  &  48 & 3.84(4) \\
  &  54 & 3.13(3) \\
  &  60 & 2.69(4) \\
  &  72 & 2.01(3) \\
\hline
4 &  32 & 3.722(17) \\
  &  36 & 2.960(16) \\
  &  40 & 2.407(15) \\
  &  48 & 1.69(10) \\
  &  52 & 1.446(8) \\
  &  56 & 1.255(12) \\
  &  64 & 0.977(12) \\
  &  80 & 0.635(13) \\
\hline
5 &  34 & 1.990(10) \\
  &  40 & 1.436(6) \\
  &  46 & 1.091(5) \\
  &  50 & 0.925(4) \\
  &  56 & 0.739(3) \\
  &  60 & 0.646(3) \\
  &  66 & 0.5352(29) \\
  & 100 & 0.2406(26) \\
\hline \hline 
\end{tabular}
\caption{The finite-volume maxima of the specific heat for various
  values of $N_s$ and $N_t$, with bootstrap errors.}
\label{table:specificheat_NsNt}
}

\subsection{Specific and latent heats}

Besides the Polyakov loop, we wish to use another thermodynamic quantity
in order to determine the order of the phase transition. We have attempted to 
measure the latent heat $L_h$ by simulating directly at the appropriate 
pseudo-critical coupling $\beta_{c,N_s,N_t}$. Based on the value of the Polyakov
loop, we have divided each ensemble into confined and deconfined 
configurations. We have then measured the action in each phase and have
determined the discontinuity. Unfortunately, this method has some difficulties 
due to the somewhat arbitrary cut in the value of the Polyakov loop used to 
distinguish confined from deconfined configurations. We have found that it is
more accurate to measure $C^{\rm max}$, the peak in the specific heat.

In Table~\ref{table:specificheat_NsNt}, we list the finite-volume
maxima of the specific heat for $N_t=3,4$, and 5. The quoted errors for
each ensemble are calculated using the bootstrap method. In
Figures~\ref{fig:c_3t}, \ref{fig:c_4t}, and \ref{fig:c_5t} we plot the
data for $C^{\rm max}/(3 N_s^2 N_t)$ for $N_t=3,4$, and 5
respectively, and for various values of $N_s$. It is no surprise that
the data do indeed extrapolate accurately in $(N_t/N_s)^2$, as a first
order transition dictates. The infinite-volume values
$C^{\rm max}/(3 N_s^2 N_t)(\infty)$ are quite small, and accurate data are 
needed in order to reach a reliable conclusion. Like the
finite-volume critical couplings, one does not know {\it a priori} how
large $N_s$ has to be before the linear regime in $(N_t/N_s)^2$ is
reached. We find that omitting the smaller $N_s$ data does not improve
the already very good linear fits, so all of the data are used in the
final analysis. The results of the fits are presented in
Table~\ref{table:specificheat}, where the errors are calculated by the
jackknife method. 

Using Equation~(\ref{eq:latent}), we convert the infinite-volume 
extrapolations into the latent heat $L_h$. We extrapolate the
dimensionless quantity $N_t^3 L_h = L_h/T_c^3$ to the continuum
linearly in $1/N_t^2$, which describes the data very well. Our
continuum determination is $L_h/T_c^3=0.188(17)$, where the error is
statistical, and the quality of the fit is $\chi^2$/d.o.f.\ =
0.90/1. With only three data points, it is not possible to estimate a
systematic error by varying the choice of fitting range. The
transition clearly becomes weaker on finer  lattices, but its first
order nature persists in the continuum limit.   

\section{Conclusions}
Our results show that $(2+1)$-d $SU(4)$ Yang-Mills theory has a first order 
deconfinement phase transition. Large and fine lattices were important for
reaching this result. In the various extrapolations and universal curves, our
assumption of using first order exponents is well confirmed by the data. The 
first order nature of the transition is further supported by the observed
coexistence of the confined and deconfined phases for all three temporal 
extensions $N_t =$ 3, 4, and 5 that we have considered. It is much harder to 
independently determine the critical exponents than to show consistency with an
expected set. Since the 2-d  $\Z(4)$-symmetric Ashkin-Teller model has 
continuously varying 
critical exponents, the challenge is particularly large. In~\cite{deF03}, by 
including logarithmic corrections to scaling, the data suggested a second order
deconfinement transition belonging to the  universality class of the 2-d 
4-state Potts model. The numerical data presented here suggest that this is not
the case. The determination of a non-zero latent heat in the limit of vanishing
lattice spacing shows that the deconfinement phase transition does not weaken
to second order but stays first order in the continuum limit.

\TABLE[]{
\begin{tabular}{ccc} 
\hline \hline
$N_t$ & $C^{\rm max}/(3N_s^2N_t)(\infty)$ & $\chi^2$/d.o.f. \\
\hline \hline
3 & $5.66(22)\times 10^{-5}$ & 5.8/8 \\
4 & $5.5(5)\times 10^{-6}$ & 0.65/6 \\
5 & $1.19(11)\times10^{-6}$ & 0.42/6 \\
\hline \hline
\end{tabular}
\caption{The extrapolated infinite-volume maxima of the specific heat,
  and the quality of the linear fits. Errors are calculated using the
  jackknife method.}
\label{table:specificheat}
}

It is surprising that so few gauge theories realize the Svetitsky-Yaffe
scenario which only applies when the deconfinement phase 
transition is second order. This may be particularly surprising in the present 
case, in which the infinite set of different universality classes of the
2-d Ashkin-Teller model would be available. In $3+1$ dimensions, 
$SU(2) = Sp(1)$ Yang-Mills theory is the only pure gauge theory with a second 
order deconfinement phase transition. In $2+1$ dimensions, the transition is 
second order only for $SU(2)$, 
$SU(3)$, and $Sp(2)$ Yang-Mills theory. Even though all symplectic groups 
$Sp(N)$ have the same center $\Z(2)$, the transition becomes first order as $N$
increases, in both $3+1$ and $2+1$ dimensions. Interestingly, the 3-d 
$\Z(N)$-symmetric spin model belongs to the universality class of the 
$U(1)$-symmetric $XY$ model for $N \ge 5$, i.e.\ the symmetry is enhanced at 
the critical point~\cite{Hov03}. However, the corresponding $(3+1)$-d $SU(N)$ 
gauge theory is unaffected by this peculiar critical behavior because its 
deconfinement transition is first order of a strength increasing with $N$. 
Indeed, as we conjectured in~\cite{Hol03}, not the center but the size of the 
gauge group determines the order of the deconfinement transition. In $3+1$
dimensions all Yang-Mills theories whose gauge group has more than three 
generators have first order transitions. As we now know, in $2+1$ dimensions 
only the Yang-Mills theories whose gauge group has at most ten generators 
(namely $SU(2)$, $SU(3)$, and 
$Sp(2)$, which has ten generators), have a second order deconfinement phase
transition. An interesting case for future study is $(2+1)$-d $G(2)$ Yang-Mills
theory. Since the exceptional group $G(2)$ has a trivial center, there is no
symmetry reason for a deconfinement phase transition and there may hence just
be a crossover. In the absence of a non-trivial center, a second order phase 
transition can be ruled out on theoretical grounds, because it would require
unnatural fine-tuning of some parameter. If the 14 deconfined $G(2)$ gluons at 
high temperature cannot smoothly crossover to the low-temperature regime 
governed by a small number of glueball states, $(2+1)$-d $G(2)$ Yang-Mills
should have a first order deconfinement phase transition. Since the 15 $SU(4)$
gluons behave in this way, based on the size of the gauge group, one may expect
the same for $G(2)$.

\acknowledgments{We wish to thank Philippe de Forcrand for useful discussions 
and Urs Wenger for providing us with his code for the re-weighting technique. 
K.H.\ is supported by the National Science Foundation under grant NSF 0704171, 
and would also like to thank the University of the Pacific for their support 
via the Eberhardt Research Fellowship. This work is also supported in part by 
the Schweizerischer Nationalfonds. The simulations were performed on a Dell 
cluster at the University of the Pacific, and on a PC cluster at the University
of Milano-Bicocca.}

\cl

\FIGURE[t]{
\vspace{2cm}
\includegraphics[width=14cm,angle=0]{mc_history_poly_60s_3t_20.40_every20th.eps}
\includegraphics[width=14cm,angle=0]{mc_history_plaq_60s_3t_20.40_every20th.eps}
\caption{\label{fig:mc_history}
The Monte Carlo histories of the Polyakov loop and the plaquette on a
$60^2 \times 3$ lattice at $\beta=20.40$, close to the deconfinement
transition. The system tunnels between the confined and four
deconfined phases, with the plaquette value tracking the change in the
Polyakov loop.} 
}

\FIGURE[t]{
\vspace{-2cm}
\includegraphics[width=11cm,angle=0]{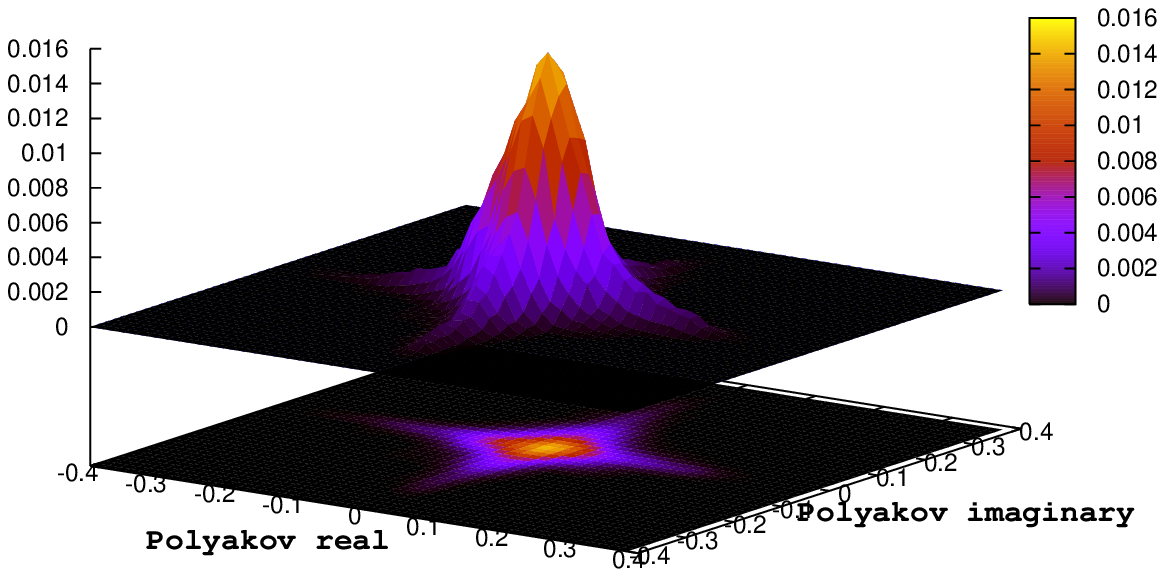} 
\vspace{-2cm}
\includegraphics[width=11cm,angle=0]{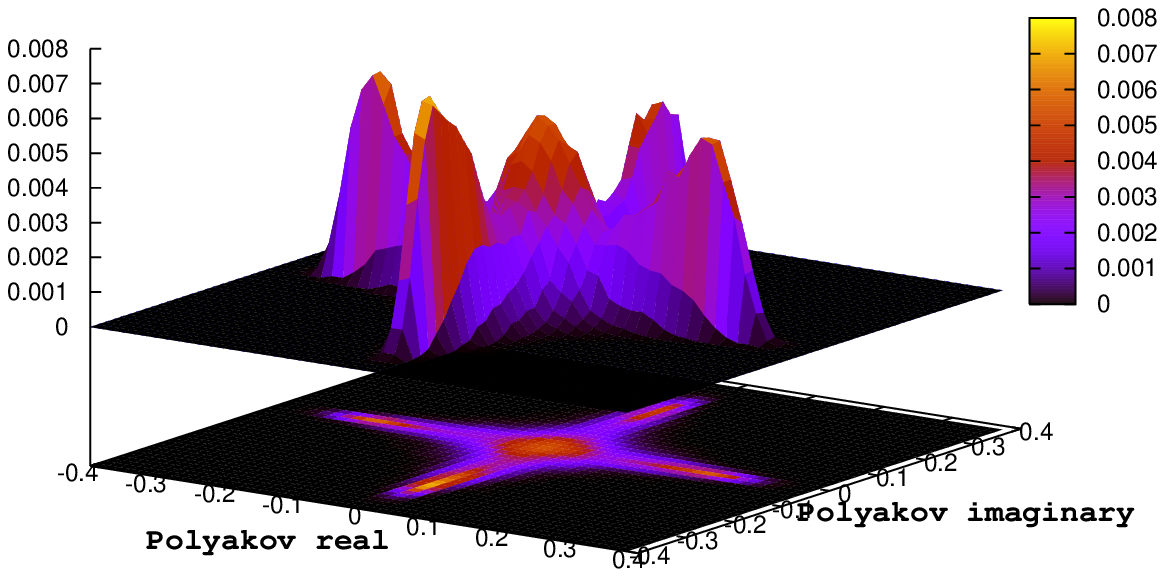}
\includegraphics[width=11cm,angle=0]{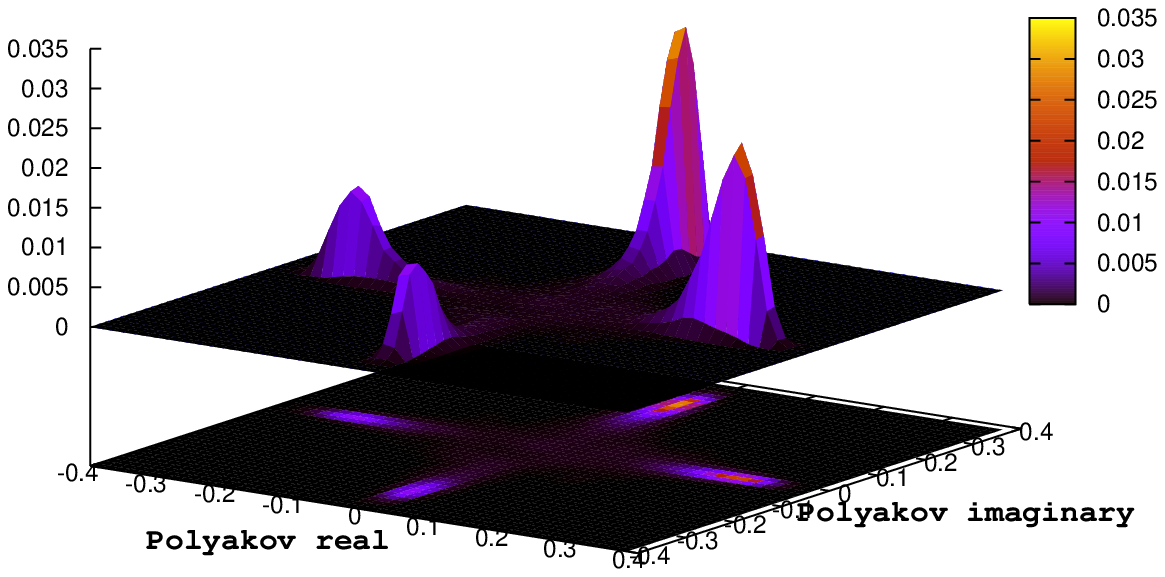}
\caption{\label{fig:poly_dist}
The probability distributions of the Polyakov loop on $36^2 \times 3$
lattices at $\beta=20.0$~(top), 20.26~(middle) and 20.5~(bottom). At
low temperature, there is a single confined phase. Close to the
critical temperature, we observe coexistence of the confined with the four
deconfined phases. At high temperature, there are only the four deconfined
phases. Because of the finite length of the simulation, the four
deconfined phases are not equally sampled.} 
}

\cl

\FIGURE[t]{
\includegraphics[width=15cm,angle=0]{distribution_plaq_36s_3t_combine.eps}
\caption{\label{fig:plaq_dist}
The plaquette value distributions for the $36^2 \times 3$ simulations shown
in Figure~\ref{fig:poly_dist}. Close to the critical temperature, at
$\beta=20.26$, we do not find two well-separated peaks, as one would
expect for a normal-strength first order phase transition.} 
}

\cl

\FIGURE[t]{
\includegraphics[width=15cm,angle=0]{sus_42s_3t.eps}
\caption{\label{fig:reweight_42s_3t}
The Polyakov loop susceptibility $\chi$ as a function of $\beta$ for
$42^2 \times 3$ lattices, using re-weighting of the combined
ensembles. The critical coupling, where $\chi$ has a maximum, is
determined to be $\beta_{c,N_s,N_t}=20.351(5)$.}  
}

\cl

\FIGURE[t]{
\includegraphics[width=15cm,angle=0]{sus_48s_4t.eps}
\caption{\label{fig:reweight_48s_4t}
The same as Figure~\ref{fig:reweight_42s_3t}, for
$48^2 \times 4$ lattices. The critical coupling is 
$\beta_{c,N_s,N_t}=26.098(10)$.} 
}

\cl

\FIGURE[t]{
\includegraphics[width=15cm,angle=0]{beta_crit_3t_fit_final_new.eps}
\caption{\label{fig:beta_crit_3t}
The critical couplings $\beta_{c,N_s,N_t}$ for $N_t=3$ and a range of $N_s$,
extracted from the peak of the Polyakov loop susceptibility $\chi$. A
linear extrapolation in $(N_t/N_s)^2$ describes the data well, as
expected for a first order transition. The infinite-volume
extrapolated value is $\beta_{c,N_t}=20.414(5)$ (jackknife error) and
the quality of the fit is $\chi^2$/d.o.f.\ = 6.9/6.} 
}

\cl

\FIGURE[t]{
\includegraphics[width=15cm,angle=0]{beta_crit_4t_fit_final_ADD_new.eps}
\caption{\label{fig:beta_crit_4t}
The same as in Figure~\ref{fig:beta_crit_3t}, this time for
$N_t=4$. The extrapolated value is $\beta_{c,N_t}=26.251(16)$ and
the quality of the fit is $\chi^2$/d.o.f.\ = 17.4/6.}  
}

\cl

\FIGURE[t]{
\includegraphics[width=15cm,angle=0]{beta_crit_5t_fit_final_new.eps}
\caption{\label{fig:beta_crit_5t}
The same as in Figures~\ref{fig:beta_crit_3t} and
\ref{fig:beta_crit_4t}, this time for $N_t=5$. For $N_t=3$ and 4, the
data show linear behavior for $(N_t/N_s)^2\le 0.015$. Hence, the two smallest 
$N_s$ values are excluded from the optimal fit. The extrapolated value is
$\beta_{c,N_t}=32.22(8)$ and the quality of the fit is $\chi^2$/d.o.f.\ = 
7.8/4.}  
}

\cl

\FIGURE[t]{
\includegraphics[width=15cm,angle=0]{universal_sus_3t_final.eps}
\caption{\label{fig:uni_sus_3t}
The rescaled Polyakov loop susceptibility $\chi/N_s^2$ versus the
finite-size scaling variable
$L^2(\beta/\beta_{c,N_s,N_t} -1)$ for $N_t=3$, where $L=N_s/N_t$. The
data fall onto a universal curve, consistent with first order exponents.} 
}

\cl

\FIGURE[t]{
\includegraphics[width=15cm,angle=0]{universal_sus_4t_final.eps}
\caption{\label{fig:uni_sus_4t}
The same as in Figure~\ref{fig:uni_sus_3t}, this time for
$N_t=4$. Again, the data are consistent with first order exponents.} 
}

\cl

\FIGURE[t]{
\includegraphics[width=15cm,angle=0]{universal_sus_5t_final.eps}
\caption{\label{fig:uni_sus_5t}
The same as in Figures~\ref{fig:uni_sus_3t} and \ref{fig:uni_sus_4t},
this time for $N_t=5$. Again, the data are consistent with first order
exponents.} 
}

\cl

\FIGURE[t]{
\includegraphics[width=15cm,angle=0]{c_3t_fit_final_new.eps}
\caption{\label{fig:c_3t}
The peak of the rescaled specific heat capacity $C^{\rm max}/(3N_s^2N_t)$ for
$N_t=3$ and a range of $N_s$. A linear fit describes the data
very well and gives the extrapolated value $C^{\rm
  max}/(3N_s^2N_t)(\infty)=5.66(22) \times 10^{-5}$. The quality of
the fit is $\chi^2$/d.o.f.\ = 5.8/8.} 
}

\cl

\FIGURE[t]{
\includegraphics[width=15cm,angle=0]{c_4t_fit_final_new.eps}
\caption{\label{fig:c_4t}
The same as in Figure~\ref{fig:c_3t}, this time for $N_t=4$. Again, we
linearly extrapolate to obtain $C^{\rm
  max}/(3N_s^2N_t)(\infty)=5.5(5) \times 10^{-6}$ and the quality of
the fit is $\chi^2$/d.o.f.\ = 0.65/6.} 
}

\cl

\FIGURE[t]{
\includegraphics[width=15cm,angle=0]{c_5t_fit_final_new.eps}
\caption{\label{fig:c_5t}
The same as in Figures~\ref{fig:c_3t} and \ref{fig:c_4t}, this time
for $N_t=5$. A linear fit of all the data gives $C^{\rm
  max}/(3N_s^2N_t)(\infty)=1.19(11) \times 10^{-6}$ and the quality
of the fit is $\chi^2$/d.o.f.\ = 0.42/6.} 
}

\cl

\FIGURE[t]{
\includegraphics[width=15cm,angle=0]{latent_heat_ntcubed_from_cmax_new.eps}
\caption{\label{fig:laten_heat_continuum}
The latent heat $L_h$, taken from the $N_s\rightarrow \infty$
extrapolation of $C^{\rm max}/(3N_s^2N_t)$. The data are extrapolated
linearly in $1/N_t^2$. The continuum value is $N_t^3 L_h=0.188(17)$
and the quality of the fit is $\chi^2$/d.o.f.\ = 0.90/1.} 
}

\cl

\end{document}